\documentclass[10pt, oneside]{article}   	
\usepackage{geometry}                		
\usepackage{graphicx}				
\usepackage{amssymb}
\usepackage{fullpage}
\usepackage{multicol}

\usepackage{hyperref}

\begin{document}
\title{   Strange Hadron Spectroscopy with the KLong Facility at Jefferson Lab 
\vspace{-6pt}}
\author{ Sean Dobbs \\ (for the KLF Collaboration)   }
\maketitle
\begin{abstract}
 The strange quark hadrons sit at an important crossroads between the light and heavy quark hadrons, but their spectrum is comparatively poorly known. The KLF experiment was recently approved to run in Hall D of Jefferson Lab, and will use an intense secondary beam of $K_L$ mesons with the existing GlueX spectrometer to collect data several orders of magnitude larger than existing dataset. In this talk, I will discuss the expected physics reach of this experiment and the status of its preparations.
\end{abstract}

 While hadron spectroscopy has undergone a renaissance in recent decades due to the availability of new large data sets and advances in our understanding of how to describe these reactions, our understanding of the spectrum of strange quark hadrons remains remarkably thin.    As an example, if we consider the spectrum of baryons containing at least one strange quark (``hyperons''), the number of ``well-established'' states (i.e. those rated 3 or 4 stars) in the Review of Particle Physics~\cite{pdg2021} we find 14~$\Lambda^\ast$ and 9 $\Sigma^\ast$ with one strange quark, 6 $\Xi^*$ with two strange quarks, and only 2 three-strange quark $\Omega^*$'s, the experimental evidence for which comes largely from kaon beam experiments in the 1970s and 80s.    This can be compared to the expectations from a recent lattice QCD (LQCD) calculation~\cite{lqcd-hyperon} of 71~$\Lambda^*$'s, 66~$\Sigma^*$'s, 73~$\Xi^*$'s and 36~$\Omega^*$'s.  Similar expectations come from constituent quark models~\cite{capstick1,capstick2}. Identifying more of these ``missing'' hyperons~\cite{missing} would allow additional, much needed, insight into baryon structure by enabling the systematic study of the baryon spectrum as a function of the number of strange quarks, and to probe the effects of QCD confinement in the transition region between the light $u$ and $d$ quarks, and the heavy $c$ and $b$ quarks.  The hyperon spectrum is also an important input into many calculations at high-baryon density and temperature, such as calculations of the neutron star equation of state~\cite{neutronstar} or the chemical potential in the very early universe~\cite{Bazavov:2020bjn}, which must currently rely heavily on models of this spectrum or the correctness of the previously mentioned LQCD calculations.  Therefore, a better understanding of the hyperon spectrum will give new insight into QCD in extreme conditions.

\begin{figure*}[!b]
	\centering
	\includegraphics[width=0.9\textwidth]{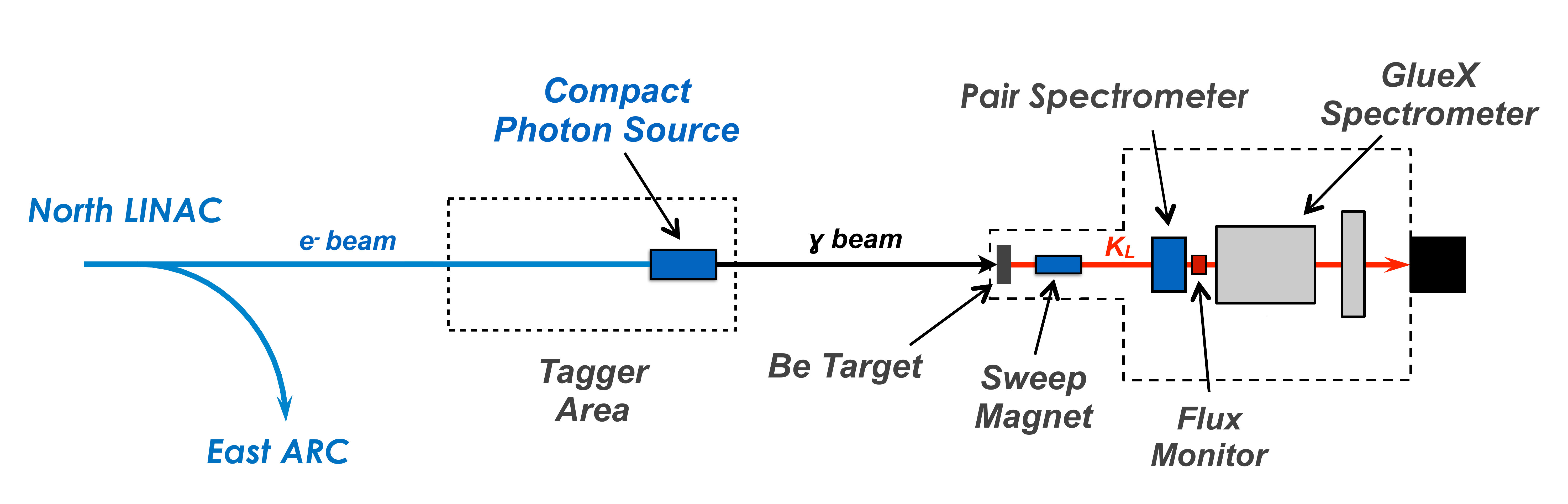}
	\caption{A diagram of the Jefferson Lab KLF, highlighting the $K_L$ beamline elements.}
	\label{fig:klf-beamline}
\end{figure*}

 The challenge facing us is to find these ``missing'' hyperons and measure their pole positions with modern analysis techniques.  A similar situation exists for the strange-quark mesons as well.  
  Recent progress has been made at photoproduction experiments like CLAS and GlueX at Jefferson Lab, and using charmed baryon decays in $e^+e^-$ annihilation experiments such as Belle~\cite{pdg2021}.  However, these measurements have been limited by the generally small cross sections of these reactions, and the limited mass range accessible in baryon decay measurements.  Kaon beam experiments naturally have much larger cross sections for strange quark hadron production, and beams of neutral kaons are particularly attractive to study due to their cleanliness and because the world data on these reactions is limited and is of insufficient precision for comprehensive resonance pole position measurements.
 To help address this situation, a new experiment at Jefferson Lab with an intense beam of $K_L$ mesons is planned to provide precise cross sections $K_L$-induced reactions and self-polarizations of hyperon decays.  This KLong Facility (KLF)~\cite{klfweb} has been approved to take data with proton and deuteron targets that will allow the identification and resonance pole position measurements of hyperons up to a mass of 2500~MeV, and allow a similar study of the kaon spectrum, including the $\kappa(800)$.  In the following, I will describe this new facility and give some highlights of the proposed experimental program.  Additional details of both can be found in Ref.~\cite{klf-proposal}.

 The planned KLF is located in Hall D at Jefferson Lab in Newport News, VA, and uses the intense, high-quality electron beam from the CEBAF accelerator to generate a tertiary beam of $K_L$ mesons.  This beam is aimed at a target of liquid hydrogen or deuterium inside the existing GlueX detector, a solenoidal spectrometer with nearly hermetic coverage for charged and neutral particles which has been described in detail elsewhere~\cite{gluex-nim}.  The primary modification to the existing Hall D configuration consists of the new $K_L$ beamline, as illustrated in Fig.~\ref{fig:klf-beamline}.  A ``continuous'' beam of 12 GeV electrons grouped in bunches separated by 64~ns with a total current of 5~$\mu$A is extracted into the hall and aimed at the Compact Photon Source (CPS)~\cite{cps}, which combines a 10\% radiation length copper radiator with a dump for the electron beam.  The photon beam thus generated is aimed at a Kaon Production Target (KPT)~\cite{kpt} containing a 40~cm Be target where the $K_L$ beam is produced, mainly through $\phi$-meson decay.  Any charged particles in the beam are swept away with a dipole magnet, and the $K_L$ flux is measured to within 5\% using $K_L$ decays in flight inside a magnetic flux monitor (KFM)~\cite{kfm}.  The experiment plans to run for 100 days on a proton target, and 100 days on a deuterium target.

\begin{figure*}[!b]
	\centering
	\includegraphics[width=0.28\textwidth]{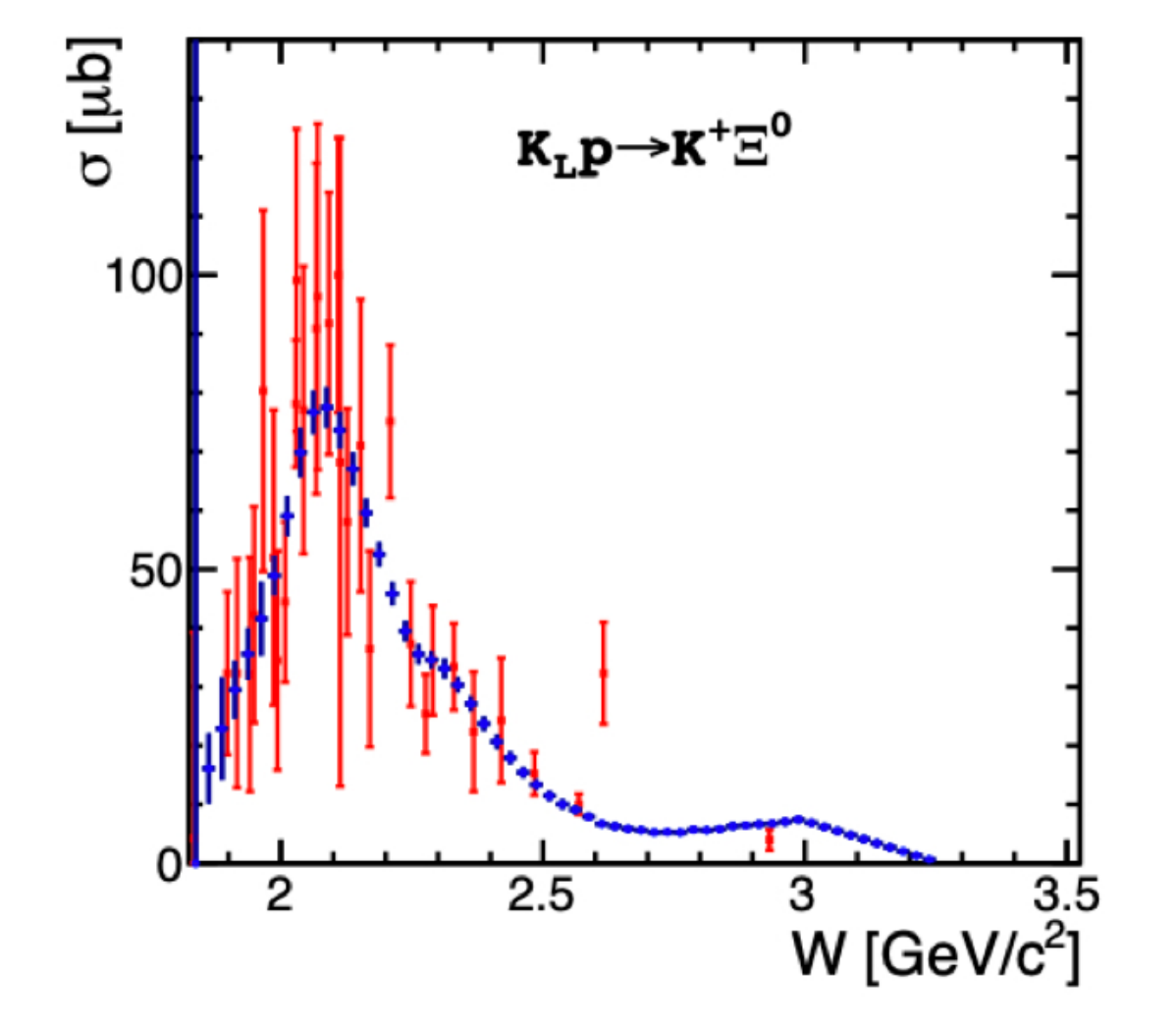}
	\includegraphics[width=0.28\textwidth]{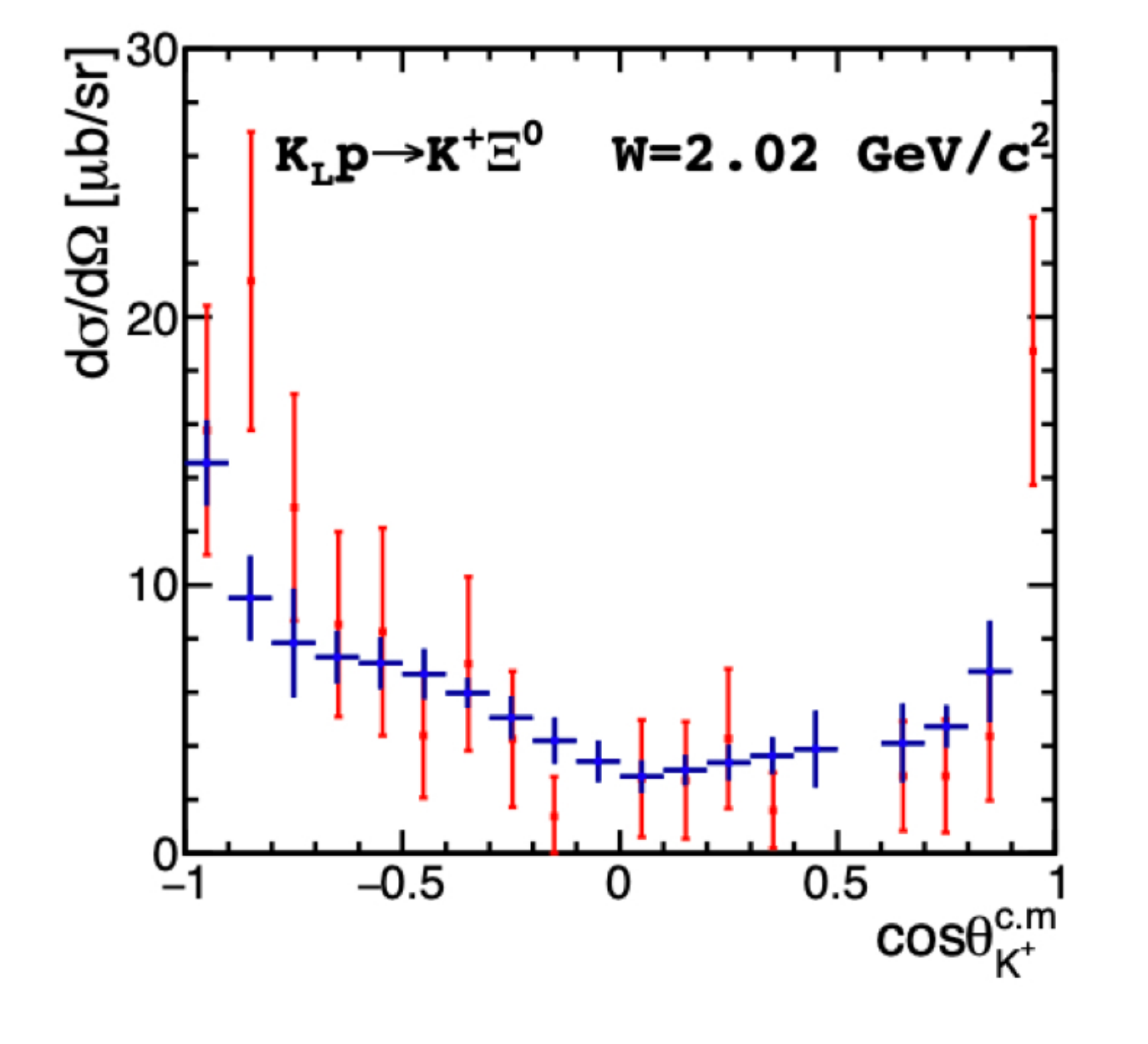}
	\includegraphics[width=0.37\textwidth]{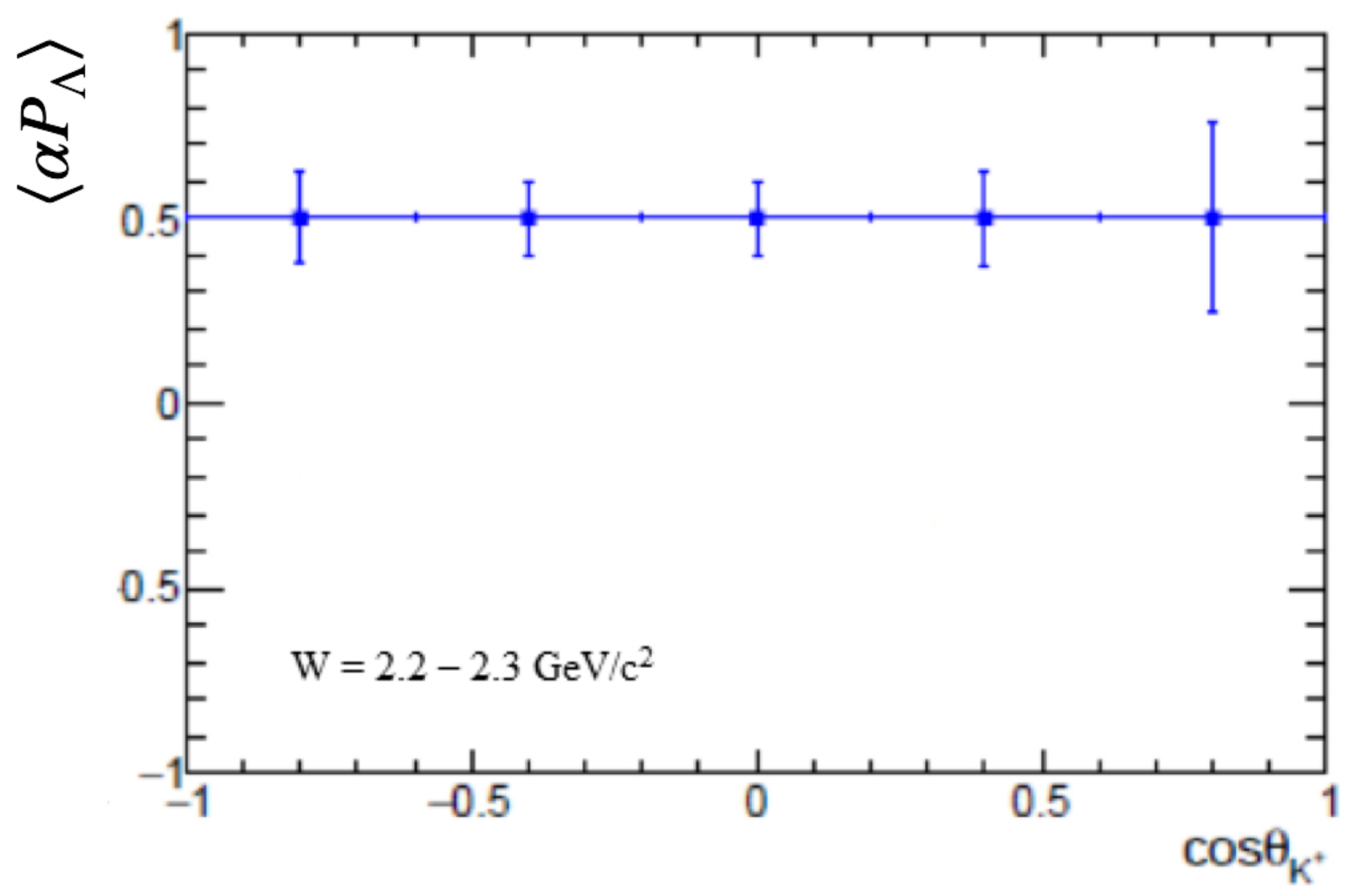}
	\caption{Results for exclusively reconstructed $K_Lp \to K^+\Xi^0$ events comparing existing data~\cite{klpxsec} (red) to projected results (blue): (Left) Total cross section as function of $W$; (Center) Cross section as a function of kaon production angle for one $W$ bin; (Right) Induced polarization of $\Xi^0$ for one $W$ bin.}
	\label{fig:klfsigma}
\end{figure*}

   The resulting $K_L$ beam provides a multi-GeV beam of $\approx10^4$~$K_L/s$, with the momentum distribution shown in Fig.~3 (left). This $K_L$ flux is $\sim1000$ times larger than that used in previous $K_L$ beam experiments in SLAC~\cite{Yamartino:1974sm}.  In Fig.~3 (left), one can also see that the neutron component of the beam mostly contributes at lower momenta where they can be separated from the $K_L$'s through time-of-flight measurements.  The $K_L$ beam energy, and therefore the center-of-mass energy of a reaction, can be well reconstructed either directly through time-of-flight measurements or indirectly through the reconstruction of all of the final state particles in the reaction.  These methods yield a good energy resolution, as shown in Fig.~2 (right).

To accomplish the goals of the hyperon part of the KLF program, differential cross sections and polarization observables must be included in a coupled-channel partial wave analysis (PWA) in order to confidently extract the hyperon spin-parity and resonance pole parameters.  We have performed detailed studies of the reconstructed of many reactions using detailed simulations of the GlueX detector, and studied the impact that these measurements would have on global PWA fits in order to determine the sensitivity of this experiment.
 
\begin{minipage}{\linewidth}
	\vspace*{10pt}
	\centering
	\includegraphics[width=0.4\linewidth]{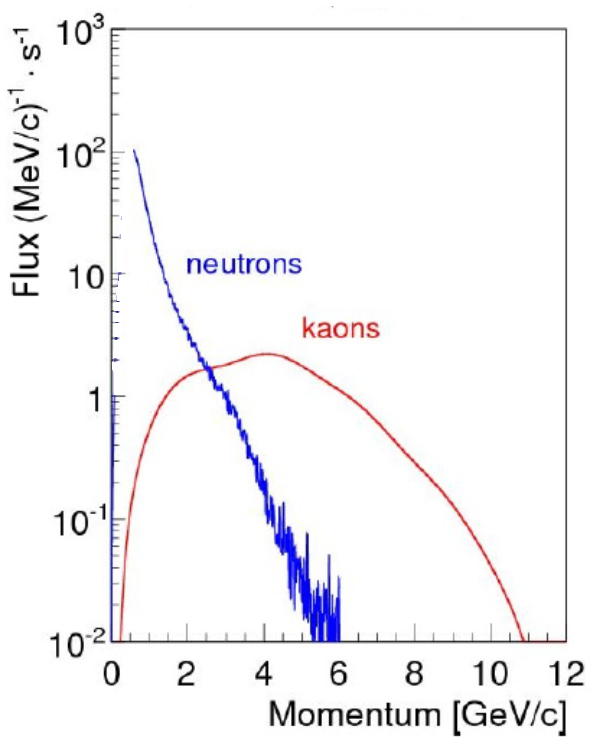}
	\raisebox{1.cm}{\includegraphics[width=0.45\linewidth]{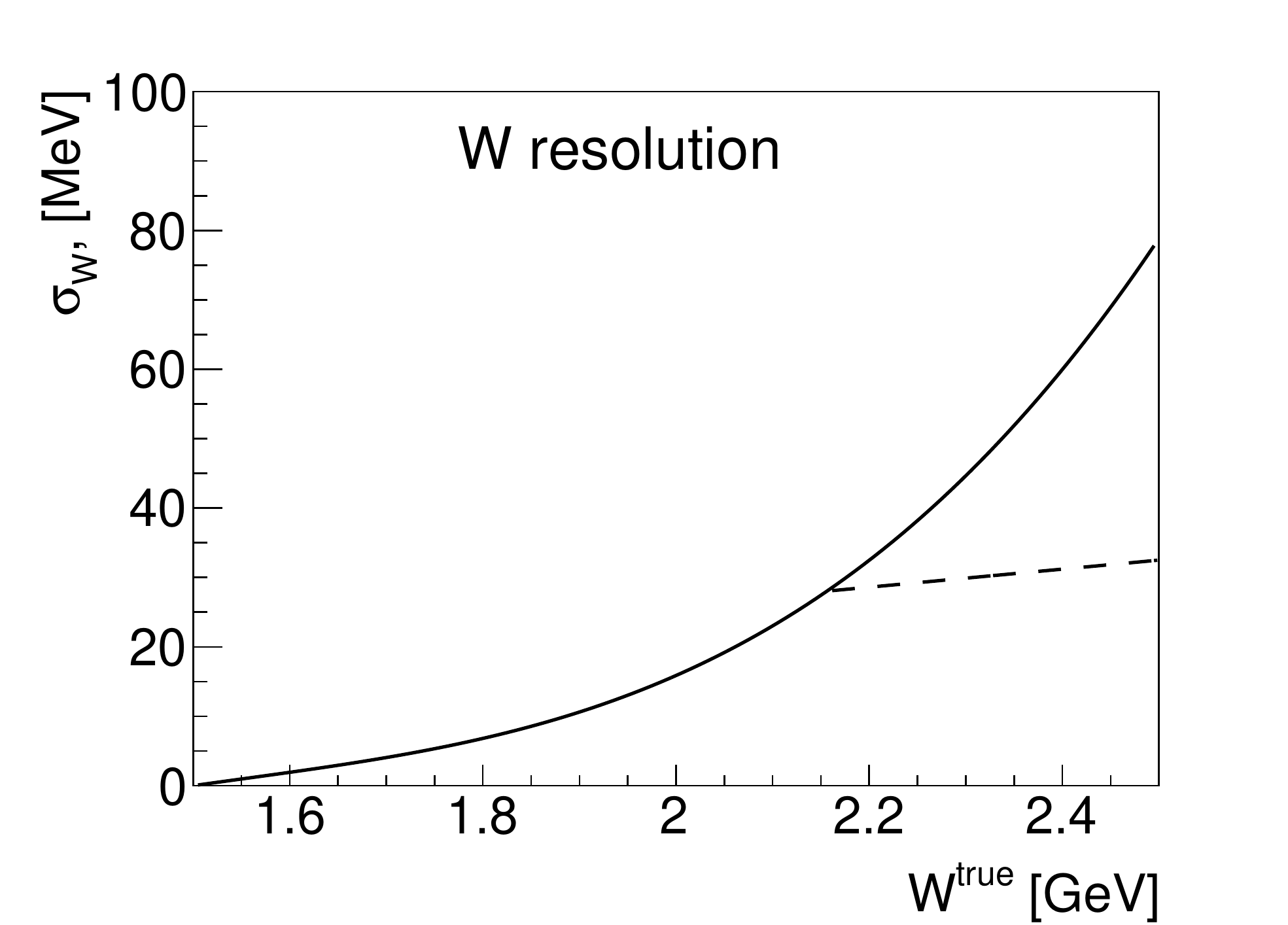}}

{\small Figure 3: (Left) Simulated kaon and neutron beam momentum spectrum on target. (Right) $W$ resolution from simulation.  The solid line gives the resolution from $K_L$ time-of-flight measurements, while the dashed line gives the expectation from exclusively reconstructed events.}

	\label{fig:klbeam}
	\vspace*{10pt}
\end{minipage}

\begin{figure*}[!tb]
\centering
{
   \includegraphics[width=0.85\textwidth,keepaspectratio]{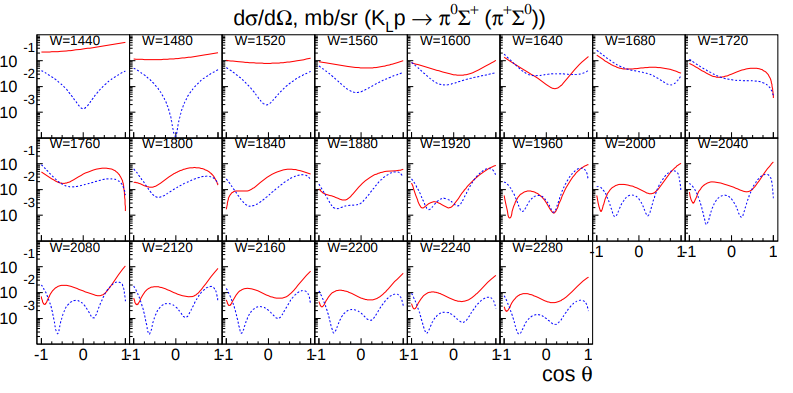} }

\setcounter{figure}{4}
\caption{ Effect of including $K_Lp$ data on three new $\Sigma^\ast$ resonances within BnGa PWA solution on differential cross sections in $K_Lp\to \pi\Sigma$ reactions~\cite{Sarantsev:2020}. The red solid line includes the 3 new resonances, the blue dashed line does not. Both models fit existing data well.}
    \label{fig:BoGa_Sigmafit}
\end{figure*}

 First, let's consider the identification of excited $\Sigma^*$ states.  If we consider two-body reactions, with a $K_L$ beam and a proton target, only $\Sigma^*$ hyperons can be produced, while both $\Lambda^*$ and $\Sigma^*$ can be produced off of a neutron.  A large range of two-body reactions can be well-reconstructed in the GlueX spectrometer, including: $K_L p \to K_Sp$, $\pi^+\Lambda$, $K^+\Xi^0$, $\pi^0\Sigma^+$, $\eta\Sigma^+$, and $\omega\Sigma^+$.  As an example of the precision of our experiment, we consider the reaction $K_L p \to K^+\Xi^0$.  In Fig.~\ref{fig:klfsigma}, we show the projected uncertainties of the total cross section, differential cross section, and self-polarization measurements for this channel.  The large improvement in statistical precision over previous measurements is clear (note that the world database of self-polarization measurements is much smaller than that of cross section measurements).  To study the precision to which pole parameters can be extracted, we simulated data for two proposed $\Sigma^*$ states:  one with $J^P=5/2^-$, $M=1.94$~GeV, and $\Gamma = 0.35$~GeV; the other with $J^P=7/2^+$, $M=1.94$~GeV, and $\Gamma = 0.4$~GeV.  The projected measurements are illustrated in Fig.~4  The uncertainties in the extracted resonance parameters are 14(36)~MeV for the mass and 14(40)~MeV for the $5/2^-$($7/2^+$) state.
  
 Another example of the importance of $K_L$ beam measurements is seen if we consider the contributions of the two isospin amplitudes $A_0$ and $A_1$ to the reaction $KN\to\Sigma\pi$:
 \begin{eqnarray}
 |A(\Sigma^+\pi^-)|^2 = \frac{1}{6}( 3|A_1|^2 + 2|A_0|^2 + 2\sqrt{6}Re(A_1A_0^*)) \\
 |A(\Sigma^-\pi^+)|^2 = \frac{1}{6}( 3|A_1|^2 + 2|A_0|^2 - 2\sqrt{6}Re(A_1A_0^*)) \\
 |A(\Sigma^+\pi^0)|^2 = \frac{1}{2} |A_1|^2
\end{eqnarray}
The $K^-p$ reactions contain contributions from $A_0$ ($\Lambda^*$-channel), $A_1$ ($\Sigma^*$-channel), and their interference.  The $K_Ln$ reactions contain the same amplitudes, but with an opposite sign for the interference term.  The $K_Lp$ reaction however only contains the $A_1$ amplitude, allowing for an unambiguous measurement.  We show the distinguishing power that this data can have by considering two different fits to the current world $\Lambda\pi/\Sigma\pi$ data:  one containing only known $\Sigma$ resonances, the other containing 3 additional expected $\Sigma$ resonances.   Both models fit existing data well.  However, the expected differential cross sections from these two models are found to differ up to two orders of magnitude, as shown in Fig.~5.  The polarization observables also show very different behavior between the two models.  This shows that our $K_L$ beam data can have a drastic impact on the search for missing $\Sigma^*$ states by resolving ambiguities in current fits to existing world data.
 
  \begin{minipage}{\linewidth}

\vspace*{10pt}

\centering
    \includegraphics[width=0.49\linewidth,keepaspectratio]{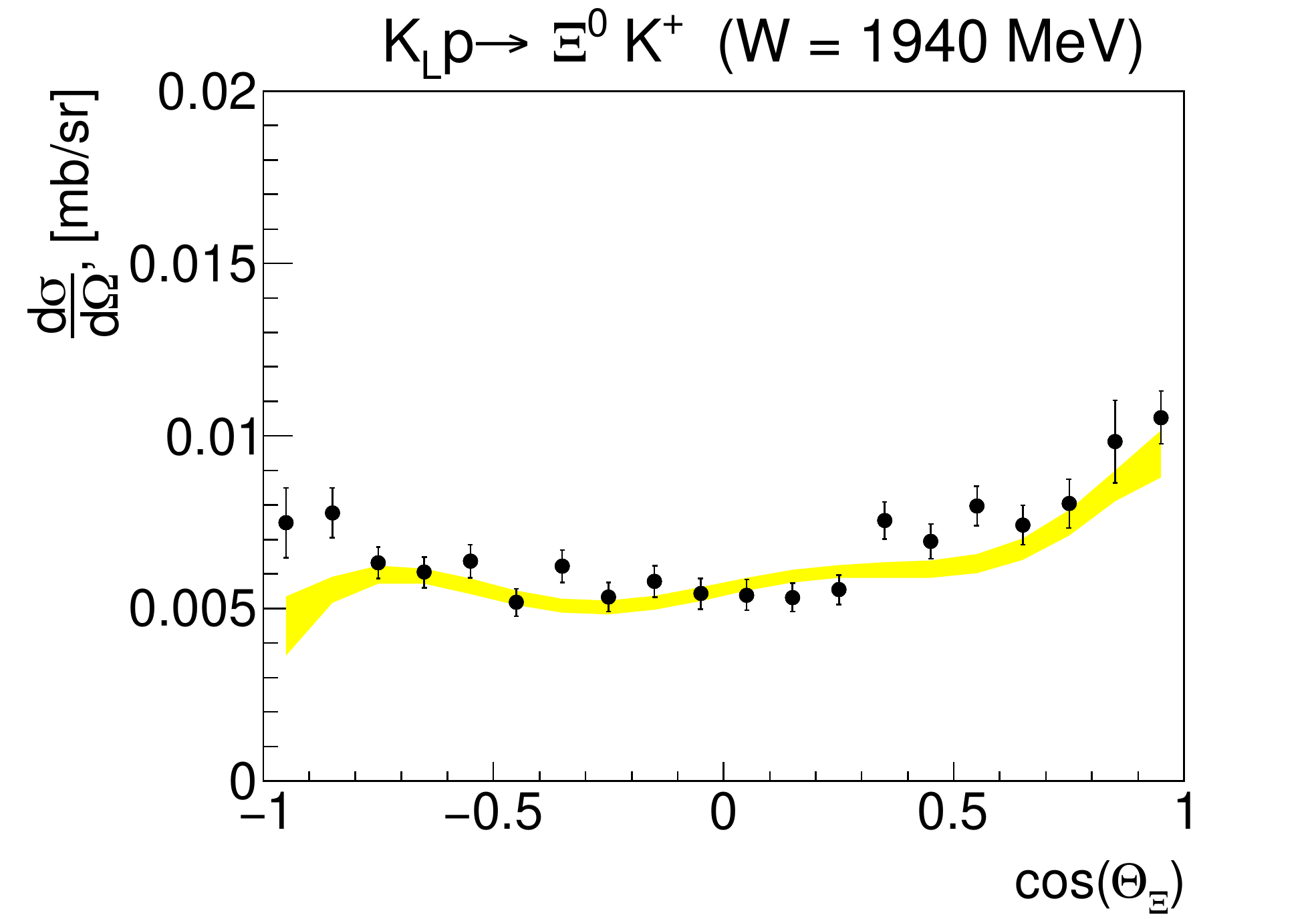}
    \includegraphics[width=0.49\linewidth,keepaspectratio]{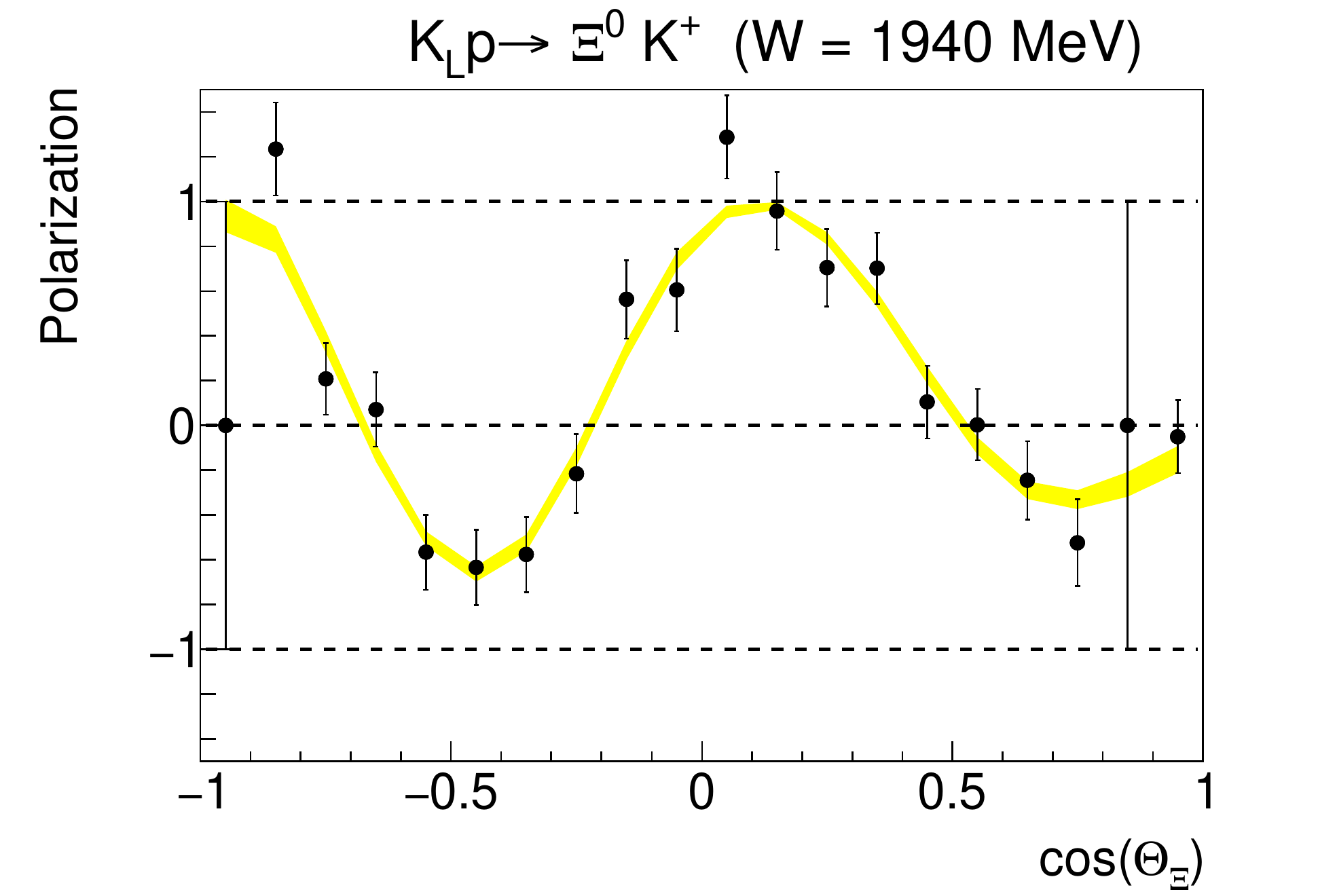}

   \includegraphics[width=0.49\linewidth,keepaspectratio]{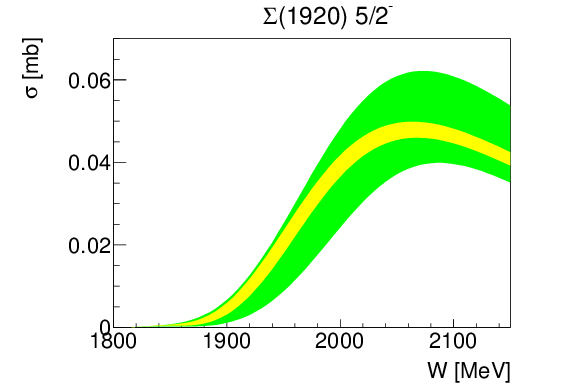}
   \includegraphics[width=0.49\linewidth,keepaspectratio]{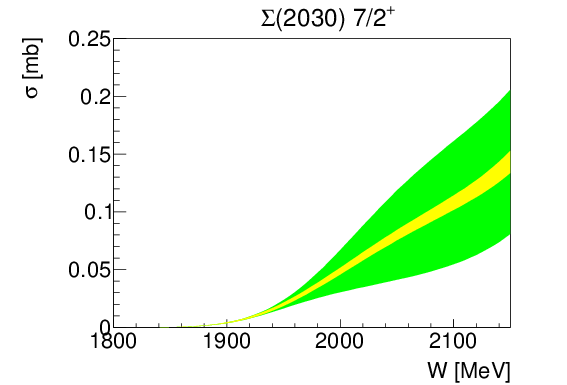} 

{\small Figure 4: Examples of the impact of the projected measurements on the BnGa PWA solutions~\cite{bngn}.  (Top) The simulated quasi-data is shown by the black points and the fit uncertainties are shown by the yellow band for $d\sigma/d\Omega$ (left) and $P$ (right) at $W = 1940~MeV$.  (Bottom) Results for $\Sigma^\ast(1920)5/2^-$ (left) and $\Sigma^\ast(2030)7/2^+$ (right) assuming 20 days of data taking (green) and 100 days (yellow).}

\vspace*{10pt}

\end{minipage}

Our data is also expected to have a large impact on the identification of $\Xi^*$ states.  Little is known of these states, with only 6 considered well-known by the PDG~\cite{pdg2021}, and almost nothing is known about their $J^P$ quantum numbers, with most of our knowledge coming from $K^-$ beam experiments in the 1960s-80s.  We can search for these excited cascade states in associated production through the decay of an intermediate hyperon state:  $K_L N \to (\Lambda^*,\Sigma^*) \to K^+ \Xi^**$.  Again the GlueX spectrometer can reconstruct many important final states well:  $\Xi^*\to \Lambda K$, $\Xi\pi$, $\Xi\eta$, $\Xi\omega$, $\Sigma K$.  Based on the previous $K^-p$ measurements, we expect large cross sections for many $\Xi^*$ states, in the $\sim1-10\mu b$ range. We have performed detailed studies on the discovery potential of KLF data, as shown in Fig.~6 for the $\Xi^*\to K\Lambda$ channel, and find that we expect to identify the known $\Xi^*$ states and likely measure their spin-parity, and have good sensitivity to identify other missing $\Xi^*$ states.

\begin{minipage}{\linewidth}
\vspace*{10pt}
	\centering
	\includegraphics[width=0.5\linewidth]{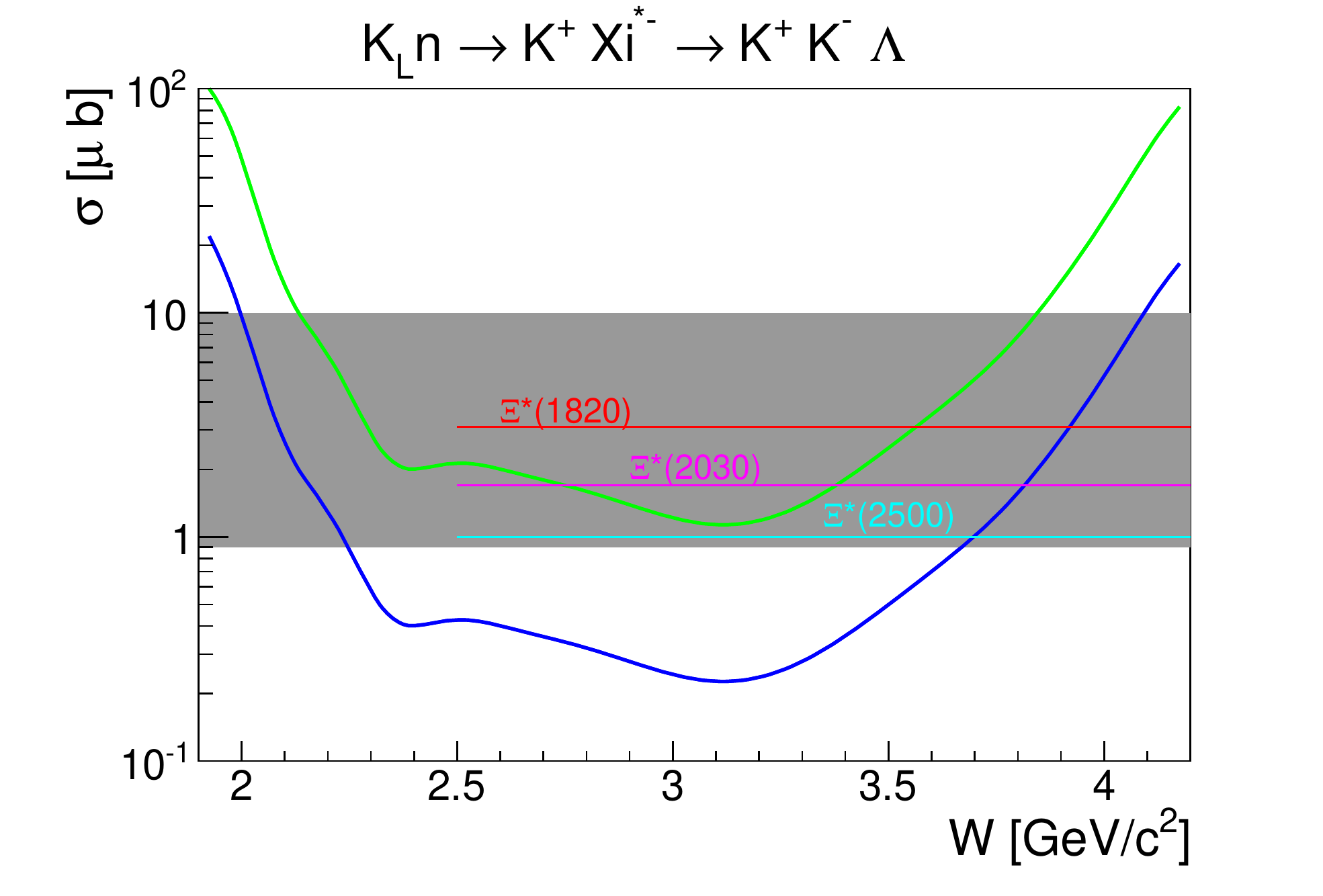}
	\includegraphics[width=0.5\linewidth]{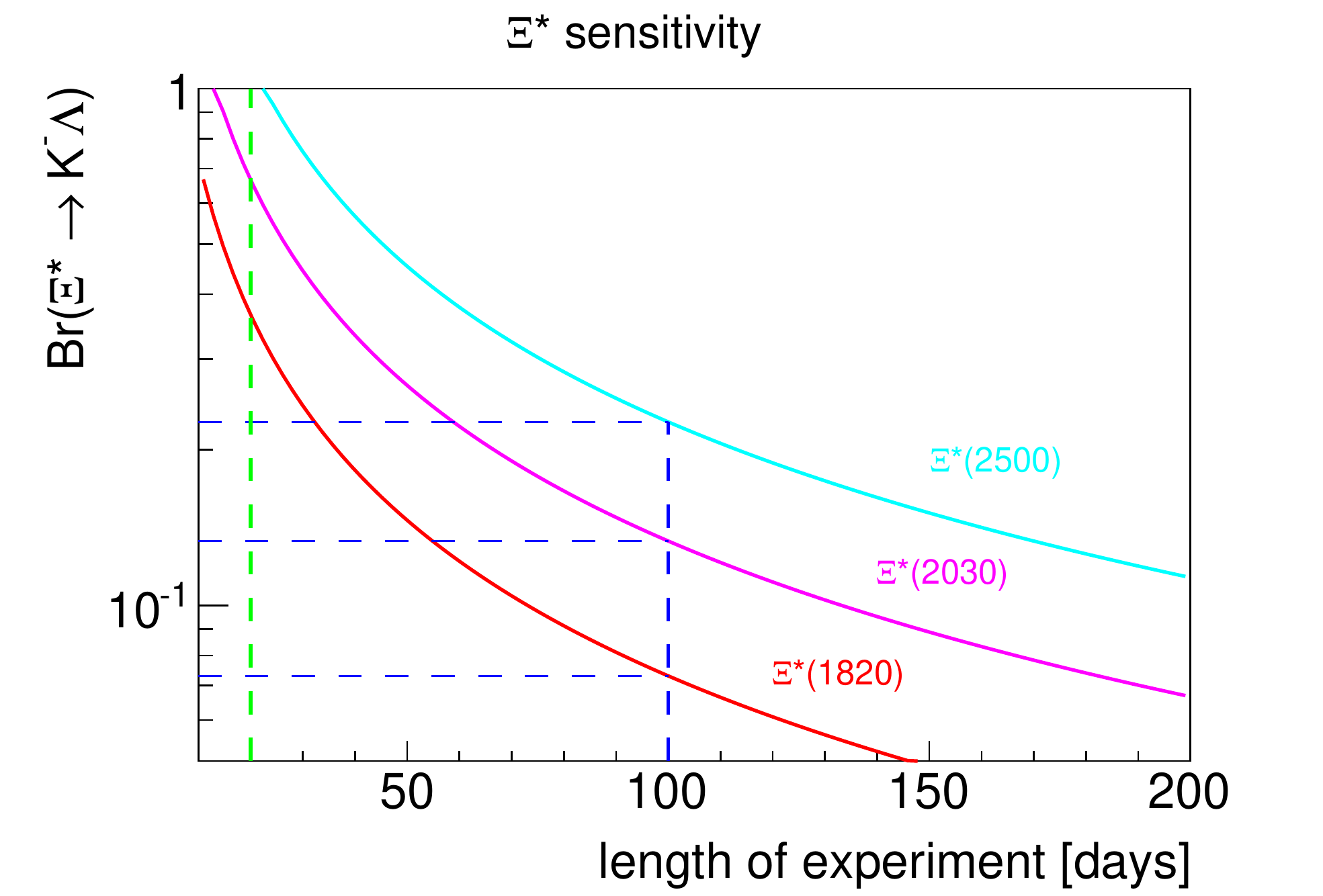}

{\small Figure 6: (Top) Projected $\Xi^*$ discovery sensitivity assuming 10~\% statistical  accuracy and $Br(\Xi^\ast\to\bar K\Lambda) = 1$ for 20 (green) and 100 (blue) days of data taking.  The gray band corresponds to typical $\Xi^\ast$ cross sections, and the red dashed lines correspond to production cross sections measured with $K^-$ beams at BNL from~Ref.~\protect\cite{Jenkins:1983pm}.
 (Bottom) Projected sensitivity to $\Xi^\ast\to\bar K\Lambda$ branching fraction as a function of experiment duration for several $\Xi^*$ states at $W = 3.1\pm0.025$~GeV. Two benchmark cases of 100 (20)~days are highlighted by dashed blue (green) curves}

\vspace*{10pt}
\end{minipage}

The KLF data is also expected to make major contributions to kaon spectroscopy and studies of $K\pi$ scattering.  Again, most of our knowledge of the kaon spectrum comes from $K^-$ beam experiments, though PWA of charmed hadron decays has provided more recent insight.  The large KLF data set (roughly 50 times the size of that collected by LASS, for example) along with modern amplitude analysis techniques will allow the study of many aspects of kaon spectroscopy, including the precision determination of the $K^*(892)$ pole parameters, the search for high-mass/high-spin kaon states, and the precision study of the scalar $K\pi$ system, which includes the ability to study in detail all four isospin partners of the lightest scalar kaon, the $\kappa/K^{*0}(700)$.  

\begin{minipage}{\linewidth}
\vspace*{10pt}
\centering
\includegraphics[width=0.5\linewidth]{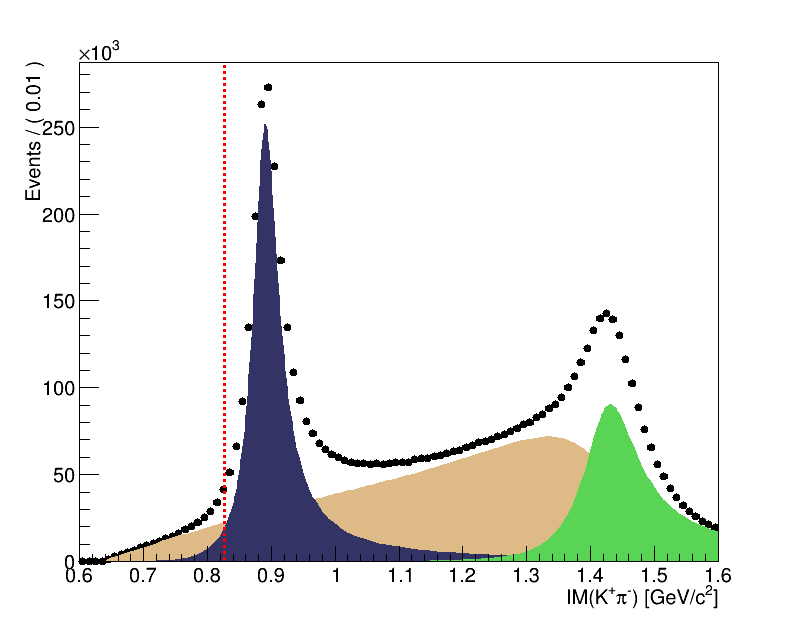}\\
\includegraphics[width=0.5\linewidth]{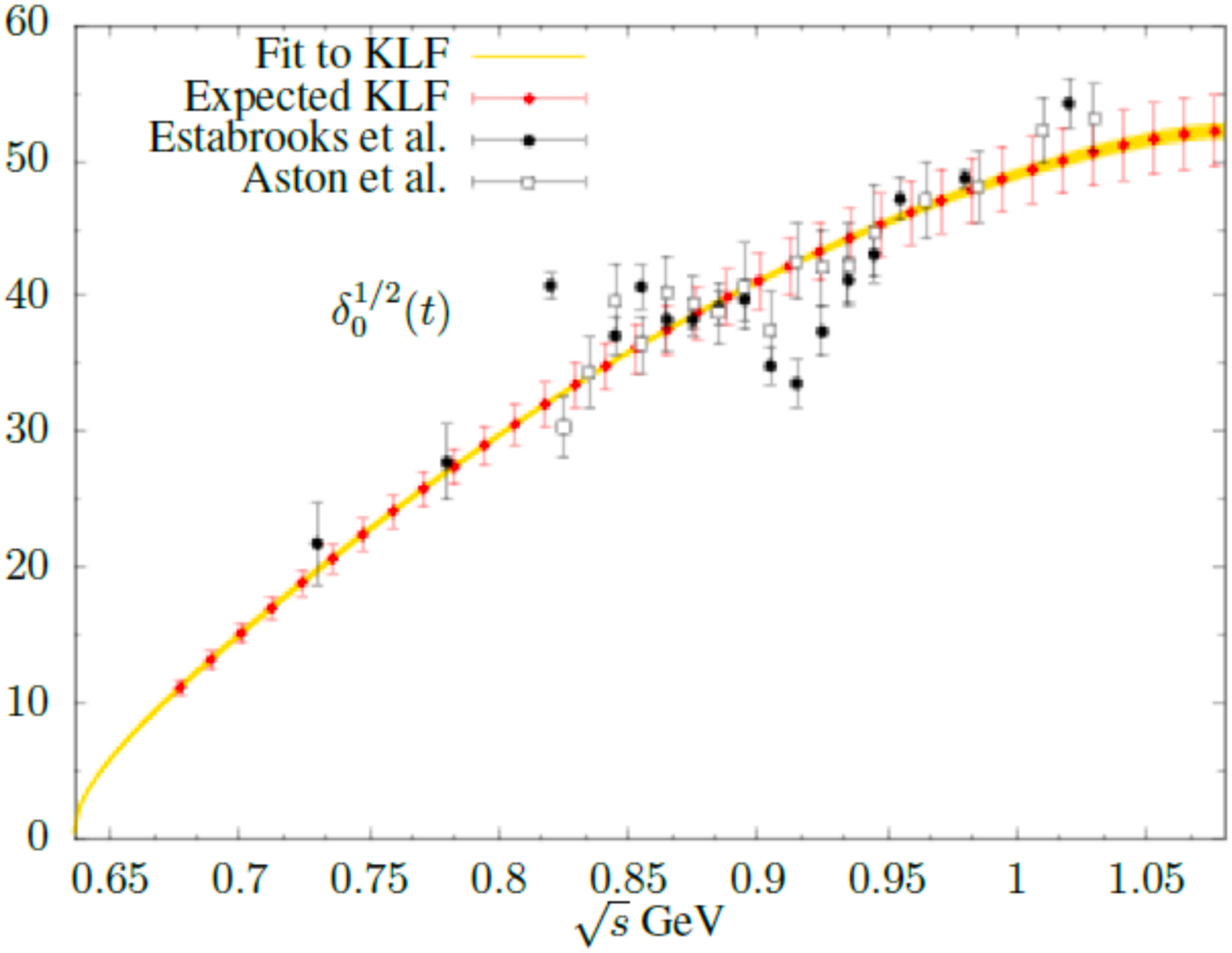}\\
\includegraphics[width=0.5\linewidth]{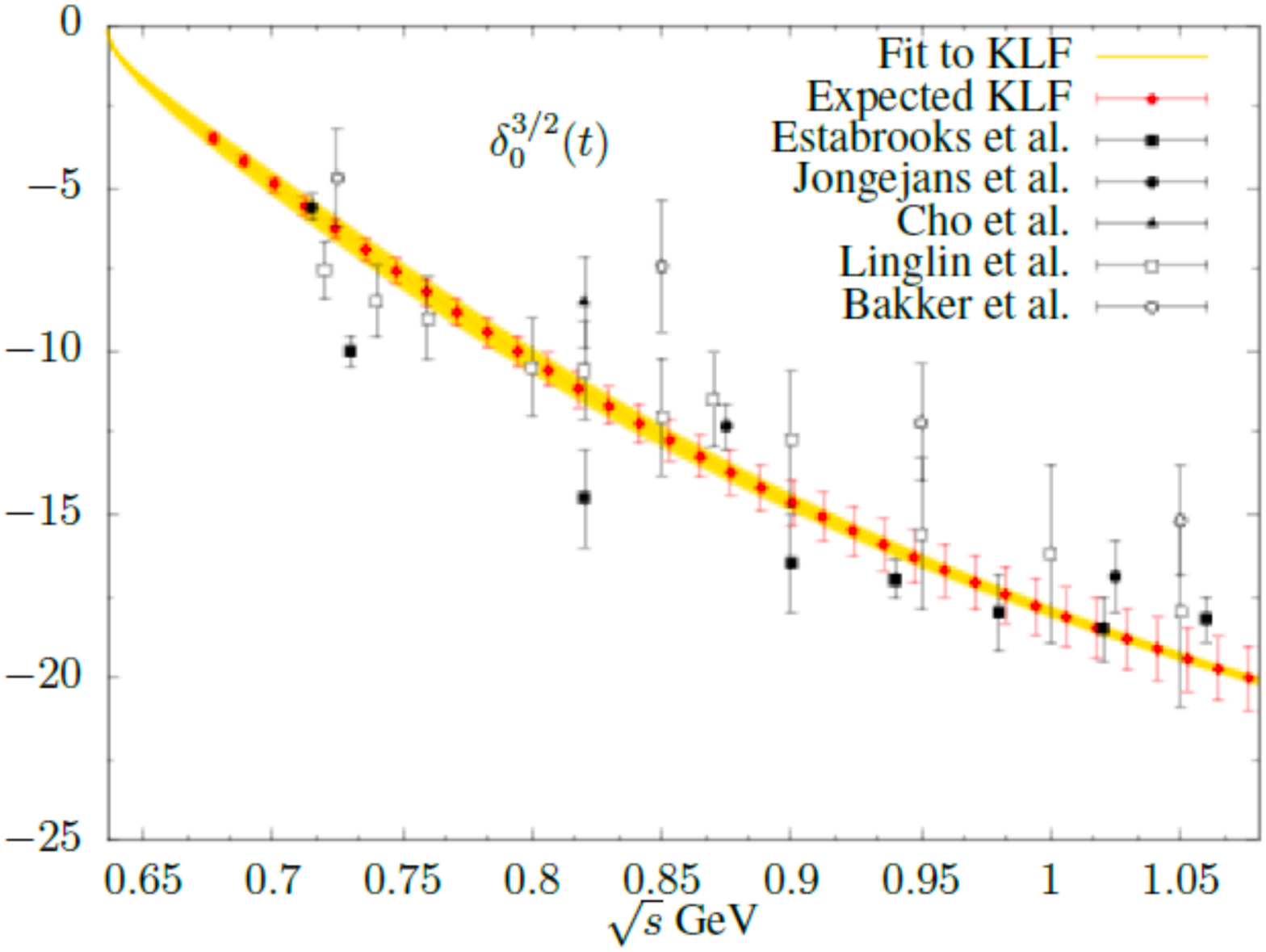}\\

{\small  Figure 7: (Top) Expected $K^+\pi^-$ invariant mass distribution with the $K^+\pi^-$ $S$-wave (light brown), $P$-wave (dark blue) and $D$-wave (green) contributions shown by the shaded curves.
The $S$-wave phase-shift for  (middle) isospin $I = 1/2$ and (top) isospin $I = 3/2$ amplitudes is given as a function of the invariant mass of the $K\pi$ system ( $\sqrt{s}$). The yellow band corresponds to the uncertainty of the fit to the current world data~\cite{kpi1,kpi2,kpi3,kpi4,kpi5,kpi6} while the red points with error bars correspond to the projected KLF measurement.}

\vspace*{10pt}
\end{minipage}

Taking the study of the scalar $K\pi$ system as an example, a major challenge is that the $K\pi$ S-wave has contributions from both the $I=1/2$ and $I=3/2$ isospin channels.  If we consider the scattering of a $K_L$ beam off a proton target, there are 4 $K\pi$ channels with a charged kaon in the final state -- $K^\pm\pi^\mp p$, $K^+\pi^0n$, $K^-\pi^0\Delta^{++}$ -- and 5 channels with a neutral kaon in the final state -- $K_L\pi^0p$, $K_{(L,S)} \pi^+n$, $K_{(L,S)} \pi^- \Delta^{++}$.  These reactions are composed of different combinations of the the $I=1/2$ and $I=3/2$ isospin amplitudes, so the challenges in describing these channels become identifying $K_L$'s and analyzing the neutral kaon final states, and developing the detailed models for all these reactions in order to fully describe the underlying $K\pi$ scattering amplitudes.  We have performed detailed simulations of several of these final states, and have also made estimates of the precision for which we can extract the S-wave phase shifts in both isospin channels, as illustrated in Fig.~7.  These studies show that we can make precise measurements over a large mass range, particularly down to low $K\pi$ masses which were not covered well by previous experiments, which are crucial for the determination of the $K^*(700)$ pole position.

To summarize, the KLF at Jefferson Lab will collect data for $K_L$ scattering off of the proton and neutron with a size and accuracy several orders of magnitude over previous experiments.  Detailed simulation studies have shown that measurements of cross sections and polarizations will allow the identification and study of hyperons up to masses of 2.5 GeV, and detailed studies of kaon spectroscopy and $K\pi$ scattering.  In addition, this unique set of data will allow for many other studies, such as neutron-induced reactions, hyperon decays, and searches for exotic hadrons.  The technical design of new hardware components and further simulation studies are ongoing.  The experimental program is expected to begin once the currently approved Hall~D photon beam program has completed, which is currently expected to run until 2025.  All are welcome who are interested in joining on this adventure!



\begin{thebibliography}{99}
%


\bibitem{pdg2021}  P.A. Zyla \textit{et al.} [Particle Data Group], \textit{The Review of Particle Physics}, Prog. Theor. Exp. Phys. 2020, 083C01 (2020) and 2021 update.

\bibitem{lqcd-hyperon} R. G. Edwards \textit{et al.} [Hadron Spectrum Collaboration], \textit{Flavor structure of the excited baryon spectra from lattice QCD}, Phys. Rev. D \textbf{87} (2013) 054506.

\bibitem{capstick1} S. Capstick and N. Isgur, \textit{Baryons in a relativized quark model with chromodynamics}, Phys. Rev. D \textbf{34} (1986) 2809.

\bibitem{capstick2} S. Capstick and W. Roberts, \textit{Quark models of baryon masses and decays}, Prog. Part. Nucl. Phys. \textbf{45} (2000) S241.

\bibitem{missing} R. Koniuk and N. Isgur, \textit{Where Have All the Resonances Gone? An Analysis of Baryon Couplings in a Quark Model With Chromodynamics}, Phys. Rev. Lett. \textbf{44} (1980) 845.

\bibitem{neutronstar} I. Vidana, \textit{Hyperons and neutron stars} Nucl. Phys. A \textbf{914}, 367 (2013);  \textit{Hyperons in Neutron Stars}, J. Phys. \textbf{668} (2016) 012031.

\bibitem{Bazavov:2020bjn}  For example, A.~Bazavov {\it et al.}, \textit{Skewness, kurtosis, and the fifth and sixth order cumulants of net baryon-number distributions from lattice QCD confront high-statistics STAR data} Phys. Rev. D \textbf{101} (2020) 074502.


\bibitem{klfweb} \texttt{https://wiki.jlab.org/klproject}

\bibitem{klf-proposal}  M. Amaryan \textit{et al.} [KLF Collaboration], \textit{Strange Hadron Spectroscopy with Secondary KL Beam in Hall D}, \texttt{arXiv:2008.08215}

\bibitem{gluex-nim} S. Adhikari [GlueX Collaboration], \textit{The GLUEX beamline and detector}, Nucl. Instr. and Meth. A \textbf{987} (2021) 164807.

\bibitem{cps} D. Day \textit{et al.},  [CPS Collaboration], \textit{A Conceptual Design Study of a Compact Photon Source (CPS) for Jefferson Lab}, Nucl. Instrum. Meth. A \textbf{957} (2020) 163429.

\bibitem{kpt} I. Strakovsky \textit{et al.}, \textit{Conceptual Design of Beryllium Target for the KLF Project}, \texttt{arXiv:2002.04442[physics.ins-det]}.

\bibitem{kfm} M. Bashkanov  \textit{et al.}, \textit{KL Flux Monitor}, KLF Note, 2018, unpublished.

\bibitem{Yamartino:1974sm} R.~Yamartino {\it et al.},  \textit{A Study of the Reactions $\overline{K}^0 p \to \Lambda \pi^+$ and  $\overline{K}^0 p \to \Sigma^0 \pi^+$ from 1~GeV/$c$ to 12~GeV/$c$} Phys.\ Rev.\ D{\bf 10} (1974) 9 ; Ph.~D Thesis, SLAC Stanford University, 1974; SLAC-R-0177, SLAC-R-177, SLAC-0177, SLAC-177.



\bibitem{klpxsec} P. Capiluppi, G. Giacomelli, G. Mandrioli, A. M. Rossi, P. Serra-Lugaresi, and L. Zitelli, \textit{A Compilation of $K_L^0p$ Cross Sections}, IFUB-81-25.

\bibitem{bngn} A. V. Sarantsev, \textit{The recent results in the analysis of the meson production reactions}, EPJ Web Conf. 199, 01009 (2019);  M. A. Matveev and A. Sarantsev, \textit{The Bonn-Gatchina analysis of the data from the $Kp$ collision reactions} PoS Hadron 2017 (2018) 069.

\bibitem{Sarantsev:2020}   A.~Sarantsev, \textit{Search for missing Sigma-hyperon states}, talk at the KLF Collaboration meeting on Feb.~12, 2020,
  \url{https://wiki.jlab.org/klproject/index.php/February_12th,_2020}.

\bibitem{Jenkins:1983pm}   C.~M.~Jenkins \textit{et al.}, \textit{Existence of $\Xi$ Resonances Above 2~GeV} Phys.\ Rev.\ Lett.\  {\bf 51} (1983) 951.

\bibitem{kpi1} P. Estabrooks \textit{et al.}, \textit{Study of $K \pi$ Scattering Using the Reactions $K^+- p \to K^\pm \pi^+ n$ and $K^\pm p \to K^\pm \pi^- \Delta^{++}$ at 13~GeV/$c$}, Nucl. Phys. B \textbf{133} (1978) 490.

\bibitem{kpi2} D. Aston \textit{et al.}, \textit{A Study of $K^- \pi^+$ Scattering in the Reaction $K^- p \to K^- \pi^+ n$ at 11~GeV/$c$}, Nucl. Phys. B \textbf{296} (1988) 493.

\bibitem{kpi3} Y. Cho \textit{et al.}, \textit{Study of $K^- \pi^-$ scattering using the reaction $K^- d \to K^- \pi^- p p_s$}, Phys. Lett. B \textbf{32} (1970) 409.

\bibitem{kpi4} A. M. Bakker \textit{et al.},  \textit{A determination of the $I=3/2$ $K\pi$ elastic-scattering cross section from the reaction $K^- n \to p K^- \pi^-$ at 3~GeV/c}, Nucl. Phys. B \textbf{24} (1970) 211.

\bibitem{kpi5} D. Linglin \textit{et al.}, \textit{$K^- \pi^-$ elastic scattering cross-section measured in 14.3~GeV/c $K^- p$ interactions}, Nucl. Phys. B \textbf{57} (1973) 64.

\bibitem{kpi6} B. Jongejans \textit{et al.}, \textit{Study of the $I = 3/2$ $K^- \pi^-$ Elastic Scattering From the Reaction $K^- p \to K^- \pi^- p \pi^+$ at 4.25~GeV/c Incident $K^-$ Momentum}, Nucl. Phys. B \textbf{67} (1973) 381.




\end{thebibliography}
\end{document}